\documentclass{revtex4-1}

\usepackage{graphics}

\begin{document}

\newcommand{\bmb}[1]{\mathbf{#1}}
\newcommand{\U}{{\sf U}}
\renewcommand{\S}{{\sf S}}
\newcommand{\bbmb}[1]{\mathbf{#1}}
\newcommand{\ms}{m^{*}}
\renewcommand{\S}{{\sf S}}
\renewcommand{\dag}{+}
\newcommand{\bbm}{\bbmb}

\title[]{Propagation of  collective surface plasmons in 1D periodic ionic structure}

\author{W. Jacak}

\affiliation{Institute of Physics, Wroc{\l}aw University of Technology,
Wyb. Wyspia{\'n}skiego 27, 50-370 Wroc{\l}aw, Poland,\\ 
 \email{witold.aleksander.jacak@pwr.edu.pl}  }

\date{Received: date / Accepted: date}

\begin{abstract}
The propagation of the collective surface plasmons, called plasmon-polaritons, in infinite equidistant  ionic sphere chain has been analyzed. The ideal cancellation of irradiative losses of these ionic excitations in the chain is demonstrated by inclusion of appropriately retarded  near-, medium and far-field components of dipole interaction between spheres in the chain. It is proved that the Lorentz friction losses in each sphere  are completely compensated by the energy income from the rest of the chain for a wide sector of the plasmon-polariton wave vector domain. There is shown  that the damping of plasmon-polariton is  reduced to only residual Ohmic losses much lower than irradiation losses for the separate electrolyte sphere.   The self-frequencies and the  group velocities  of plasmon-polaritons for longitudinal and transversal (with respect to   the chain orientation)  polarizations are determined and assessed for various ion and electrolyte parameters.  It is proved that there exist weakly damped self-modes of plasmon-polaritons in the chain for which  the propagation range is limited by relatively small Ohmic losses only. Completely undamped collective waves are also described in the case of the 
presence of persistent external excitation of some fragment of the chain. 
The possibility of application of the plasmon-polariton model to describe the so-called saltatory conduction in periodically myelinated nerve axons is preliminarily discussed.
\end{abstract}


\maketitle
\section{Introduction}

In the previous paper \cite{jacak2014a} we have developed the description of plasmon scillations in ionic spheres in analogy to plasmons in metallic nanospheres. We have demonstrated the existence of the surface and the volume modes of plasmons in electrolyte spheres  and  analyzed their attenuation including scattering and irradiation lesses, the later expressed by the Lorentz friction of oscillating charge fluctuations. Majority of details for ionic systems remain in analogy to electrons in metals though with overall scale shifted $3-4$ orders in magnitudes   
toward lower frequencies, larger system size, and larger wave lengths of resonances. 

In the present paper we  will utilize plasmonic properties of ionic electrolyte  spheres in order to discuss kinetics of effective surface plasmon wave-type excitations in a chain constructed of electrolyte spheres, in analogy to plasmon-polaritons propagating in metallic nano-chains.  
A question which we try to discuss in the present paper is the possibility of similar plasmon-polariton  properties occurrence for ionic carriers instead of electrons and the verification of efficiency of electrolyte plasmon wave-guides. The electrolyte spheres of the chain  are assumed to be   closed by sphere-shaped membranes similar as those  frequent in   biological structures.
 The utilizing of ions in cell signaling, membrane transfer or nerve cell conductivity  are the examples of functioning of such ionic structures and the question on the soft plasmonics usefulness in there arises. 

For the theoretical model we will  consider an infinite equidistant chain of spheres with radius $a$, though a generalization to oblate or prolate spheroids or even other shapes of the chain elements is natural. 
The ionic systems in each electrolyte sphere we will model upon the developed in the previous paper \cite{jacak2014a} the simplified effective jellium approach for  ions.
Taking in mind that  metallic nanochains create very effective low-noise wave-guides for electromagnetic signals in  form of the wave-type  plasmon-polaritons,  we will  model the similar phenomenon in ionic sphere chains.

\begin{figure}[h]
\centering
\scalebox{0.45}{\includegraphics{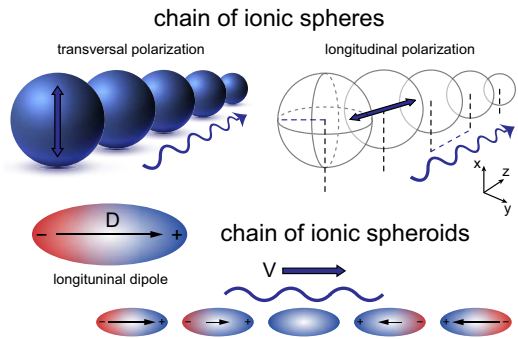}}
\caption{\label{fig1} Schematic presentation of the chain of ionic spheres with propagating plasmon-polariton (upper);  the dipole $\mathbf{D}$ creation in single spheroid and wave type dipole excitation in the chain of spheroids (the longitudinal mode is depicted in the prolate geometry) traversing the chain  with the group velocity $\textbf{v}$  (lower) }
\end{figure}

In the next section of the present paper we will derive the fundamental equation for dynamics of collective surface plasmons in the ionic sphere chain including the surface plasmon self-oscillations in each chain component, the Joule heat damping of plasmons due to irreversible energy dissipation (Ohmic losses) by microscopic scattering of ions, the irradiation losses described by the Loretz friction hampering the charge movement and the energy mutual supplementation in the chain via radiation of all elements of the chain. The balance of the energy will be analyzed in order to detect low-noise windows for ionic plasmon-polariton propagation. The solution of dynamic equation in the homogeneous and inhomogeneous cases will be presented. All the detailed model calculations will be shifted to Appendices. In the last but one Appendix the precise analysis of some singularities occurring in the dynamical equation for plasmon-polariton in the chain and caused by the medium- and far-field components of dipole inter-sphere interaction in the chain will be presented in analogy to similar features in metallic nanochains. In the last Appendix  some links to electric signaling in nerve axons are preliminary discussed  from point of view of the possible application of the developed description of ionic plasmon-polaritons in ionic periodic systems. The explanation of so-called saltatory conduction in myleanated nerve axons is suggested by utilizing of plasmon-polaritons propagation in periodic linear ionic structure. 

\section{Dynamic equation for collective surface plasmons in a chain of  spherical ionic systems}

Now we will consider an infinite  chain of equidistantly placed ionic spherical systems with radius $a$ separated with the distance $d$ between the sphere centers aligned along the axis $z$. The question is to describe collective surface dipole  plasmons in such system in analogy to propagation of plasmon-polaritons in metallic nano-chains \cite{deabajo,citrin2005}. Because of similarity of ionic plasmon excitations in finite electrolyte systems to plasmons in metallic nanosphere, one can expect also a similarity in collective plasmon behavior in complex ionis systems to metallic ordered nano-structures. Even though in finite electrolyte systems plasmon oscillations have much lower energy in comparison to plasmons in metals, their irradiation properties with typical size-dependence may admit  an effective plasmon energy transfer along the ionic chain in analogy to metallic nano-chains. Such electrolyte  chains would serve as the wave-guides for plasmon signals along ionic structures with potential link to some biological ionic information channels. 

The main property of metallic plasmonic wave-guides is their perfect transmittance, i.e., the absence of irradiation losses and the reducing of signal damping to only Ohmic attenuation, which makes possible a long range almost undamped propagation. The second property is the reduction of signal group velocity (in metallic chains) to at least one order lower value in comparison to light velocity allowing  subdiffraction arrangement of plasmon-polariton circuits. Though with shorter wave-length collective plasmon excitations in the chain in comparison to the same energy photons, the former  would efficiently transport information signals over large distances, with transfer parameters controlled in wide range by chain size and geometry. If these properties could be repeated in ionic periodic structures (with the  appropriately shifted scale of corresponding quantities), they would be of some importance for understanding of signal transfer in biological systems.

Periodicity of the chain makes the system similar to a 1D crystal. The interaction between chain elements  can be considered as of dipole-type coupling when surface dipole plasmons are excited in them. For ionic sphere chain one can adopt the result from metallic chain analysis, which justify the dipole model of sphere interactions when the distance between neighboring spheres is larger than the sphere radius, i.e., when $d>3a$ (otherwise the multipole channels may contribute inter-sphere interaction)
\cite{citrin2005,schatz2003,schatz2004}. One can   notice also, that the model of interacting dipoles \cite{bohren1987,schatz1999} was developed originally for description of stellar matter \cite{drain1988,purcell1973} and next it has been adopted to metal particle systems \cite{drain1994,markel1993}.  

The dipole interaction resolves itself to electric and magnetic field created by oscillating dipole  $\mathbf{D}(\mathbf{r},t)$ in any distant point. If this this point is marked by the vector  $\mathbf{r}_0$ (fixed to the end of $\mathbf{r}$, where the dipole is placed), the electric field  produced by the dipole   $\mathbf{D}(\mathbf{r},t)$ has the following form, including
 relativistic retardation \cite{lan,jackson1998}:
\begin{equation}
\label{electric}
\begin{array}{l}
 \mathbf{E} (\mathbf{r},\mathbf{r_0},t)=\frac{1}{\varepsilon}
\left(-\frac{\partial^2}{v^2\partial t^2} \frac{1}{r_0} -\frac{\partial}{v \partial t} \frac{1}{r_0^2}-\frac{1}{r_0^3}\right)\mathbf{D}(\mathbf{r},t-r_0/v)\\ 
+ \frac{1}{\varepsilon}
\left( \frac{\partial^2}{v^2\partial t^2}\frac{1}{r_0}+\frac{\partial }{v \partial t} \frac{3}{r_0^2}+\frac{3}{r_0^3}\right) \mathbf{n}_0(\mathbf{n}_0\cdot \mathbf{D}(\mathbf{r}, t-r_0/v)),\\
\end{array}  
\end{equation}   
where, 
$\bmb{n}_0=\frac{\bmb{r}_0}{r_0}$ and   
$v=\frac{c}{\sqrt{\varepsilon}}$, $\varepsilon$ is the dielectric susceptibility of the medium surrounding the chain.
The terms with the denominators  $r_0^3$, $r_0^2$ and $r_0$ are usually called near-field, medium-field and far-field interaction components, respectively.
The above formula allows for description of mutual interaction of plasmon dipoles on each sphere in the chain. 
The spheres in the chain are numbered by integer $l$ and 
 the equation for the surface plasmon oscillation of the $l^{{th}}$ sphere can be written as follows,
\begin{equation}
\label{tra111111}
\begin{array}{l}
 \left[\frac{\partial^2}{\partial t^2}+  \frac{2}{\tau_0}  \frac{\partial}{\partial t} +\omega_1^2\right]D_{\alpha}(ld,t)\\
 =\varepsilon \omega_1^2a^3
  \sum\limits_{m=-\infty, \;m\ne l }^{m=\infty} 
E_{\alpha}\left(md,t-\frac{|l-m|d}{c}\right)\\
+\varepsilon \omega_1^2a^3 E_{L\alpha}(ld,t) +\varepsilon \omega_1^2a^3 E_{\alpha}(ld,t).\\
  \end{array}
\end{equation}
 The first term of the r.h.s. in Eq. (\ref{tra111111}) describes the dipole  coupling  between spheres  and the other two terms correspond  to plasmon attenuation due to the Lorentz friction (as described in the previous paper \cite{jacak2014a} addressed to the single ionic sphere)  and the forcing field due to  an external electric field, correspondingly, $\omega_1=\frac{omega_p}{\sqrt{3\varepsilon}}$ is the Mie frequency of dipole surface plasmons \cite{jacak2014a}. The Ohmic losses are included by the term \cite{atwater2003},
\begin{equation}
\label{form}
\frac{1}{\tau_0}=\frac{v}{2\lambda_B}+\frac{Cv}{2a},
\end{equation}
 where $\lambda_B$ is the mean free path of carriers in bulk electrolyte, $v$ is the mean velocity of carries at the temperature $T$, $v=\sqrt{\frac{3kT}{m}}$, $m$ is the mass of the effective ion, $k$ is the Boltzmann constant, $C$ is the constant of unity order, $a$ is the sphere radius. The first term in the expression for $\frac{1}{\tau_0}$ approximates ion scattering losses, the same as in bulk electrolyte, whereas the second term displays losses due to scattering of ions on boundary of the sphere with the radius $a$. The index $\alpha$ indicates polarization, the longitudinal one, when  $\alpha=z$ and the transversal one, when  $\alpha=x(y)$ (the chain orientation is assumed in $z$ direction). According to Eq. (\ref{electric}) we get in Eq. (\ref{tra111111}),
\begin{equation}
\begin{array}{ll}
E_z(md,t)&=\frac{2}{\varepsilon d^3}
\left(\frac{1}{|m-l|^3}+\frac{d}{v|m-l|^2}\frac{\partial}{\partial t}\right)\\
&\times D_z(md,t-|m-l|d/v),\\
E_{x(y)}(md,t)&=-\frac{1}{\varepsilon d^3}\left(\frac{1}{|m-l|^3}+\frac{d}{v|m-l|^2}\frac{\partial}{\partial t}+
\frac{d^2}{v^2 |l-d|}\frac{\partial^2}{\partial t^2} \right)\\
&\times D_{x(y)}(md, t-|m-l|d/v).\\
\end{array}
\end{equation}

 Due to  periodicity of the chain  (in analogy 1D  crystal) one can assume the wave type collective solution of the dynamical equation (\ref{tra111111}),
\begin{equation}
\label{eq8}
\begin{array}{l}
D_{\alpha}\left(ld,t\right)=D_{\alpha}\left(k,t\right)e^{-ikld},\\
0\leq k \leq\frac{2\pi}{d}.
\end{array}
\end{equation}
This corresponds to the  form of a solution in  the Fourier picture of Eq.      
(\ref{tra111111}) (the discrete Fourier transform (DFT) with respect to the positions and the  continuous Fourier transform (CFT) with respect to time) similar to that one  as used in the case of phonons in 1D crystal. 
Let us note that the  
DFT is defined for a finite set of numbers, therefore we can consider the chain with $2N+1$ spheres, i.e., the chain of finite length $L= 2Nd$. Then, for any discrete characteristics $f(l),\;\;l=-N,...,0,...,N$ of the chain, like a selected polarization of dipole distribution, one deals with DFT picture $f(k)=\sum\limits_{l=-N}^{N}f(l)e^{ikld}$, where $k=\frac{2\pi}{2Nd}n,\;n=0,...,2N$. This means that $kd\in[0,2\pi)$ due to periodicity of the equidistant chain. On the whole system the Born-Karman boundary condition is imposed resulting in the form of $k$ as given above. For the infinite length of the chain one can take finally the limit $N\rightarrow \infty $ which causes that the variable $k$ is quasi-continuous, but still $kd\in[0,2\pi)$.    

In the manner which is described in details  in Appendix \ref{a}, we arrive at the Fourier representation  of Eq. (\ref{tra111111}) in the following form,
\begin{equation}
\label{aaa1}
\begin{array}{l}
\left(-\omega^2-i\frac{2}{\tau_0}\omega +\omega^2_1\right)D_{\alpha}(k,\omega)\\
=\omega_1^2\frac{a^3}{d^3}F_{\alpha}(k,\omega)D_{\alpha}(k,\omega)+
\varepsilon a^3 \omega_1^2 E_{0\alpha}(k,\omega),\\
\end{array}
\end{equation}
with 
\begin{equation}
\label{aaapodluzneipoprzeczne}
\begin{array}{l}
F_z(k,\omega)=4\sum\limits_{m=1}^\infty \left(\frac{cos(mkd)}{m^3}cos(m\omega d/v)\right.\\
\left. +\omega d /v \frac{cos(mkd)}{m^2}sin(m\omega d/v)\right)\\
+2i \left[\frac{1}{3}(\omega d /v)^3+2\sum\limits_{m=1}^\infty \left(\frac{cos(mkd)}{m^3}sin(m\omega d/v)\right.\right.\\
\left.\left. -\omega d/v\frac{cos(mkd)}{m^2}cos(m\omega d/v)\right)\right],\\
F_{x(y)}(k,\omega)=-2\sum\limits_{m=1}^\infty \left(\frac{cos(mkd)}{m^3}cos(m\omega d/v)\right.\\
\left.+\omega d /v \frac{cos(mkd)}{m^2}sin(m\omega d/v)
 -(\omega d/v)^2\frac{cos(mkd)}{m}cos(m\omega d/v)\right)\\
-i \left[-\frac{2}{3}(\omega d /v)^3+2\sum\limits_{m=1}^\infty \left(\frac{cos(mkd)}{m^3}sin(m\omega d/v)\right.\right.\\
\left.\left.+\omega d/v\frac{cos(mkd)}{m^2}cos(m\omega d/v)
-(\omega d/v)^2\frac{cos(mkd)}{m}sin(m\omega d/v)\right)\right].\\
\end{array}
\end{equation}

One can analytically calculate the sums in functions $ImF_z(k,\omega)$ and $Im F_{x(y)}(k,\omega )$ corresponding to the radiative damping for longitudinal and transversal plasmon-polariton  modes, when within the perturbative approach one takes $\omega=\omega_1$ in these functions, as  is done in  Appendix \ref{a}---Eqs (\ref{podluzne111}) and (\ref{poprzeczne111}). Both these functions perfectly vanish when $0<kd\pm \omega d/v<0$ (the corresponding region is indicated in Fig. \ref{figk2}). Outside this region radiative damping  is not zero, which for longitudinal and transversal modes is illustrated in Figs \ref{figk1} and \ref{figk3}, respectively.

\section{Plasmon-polariton self-modes in the chain and propagation of e-m signal along periodic ionic structure}

The real part of the functions $F_{\alpha}$ renormalizes the  self-frequency of plasmon-polaritons in the chain, while its imaginary part renormalizes damping rate of these modes. $ReF_{\alpha}(k,\omega)$ and $ImF_{\alpha}(k,\omega)$ are functions of $k$ and $\omega$. Within the first order approximation one can put 
$\omega=\omega_1$ in $ReF_{\alpha}$ and also in residual nonzero $ImF_{\alpha}$, in the latter function, outside the region  $0<kd\pm \omega_1 d/v<2\pi$. Let us emphasize, however, that vanishing of $ImF_{\alpha}(k,\omega)$ inside the region $0<kd\pm \omega d/v<2\pi $ holds for any value of $\omega$ \cite{jacak2013} (thus also  for  exact solution for frequency and not only for approximated $\omega=\omega_1$).

Hence, within the perturbative approach  one can rewrite the dynamic equation  (\ref{aaa1}) for plasmon-polariton modes in the chain in the following form,
\begin{equation}
\label{aaa111}
\begin{array}{l}
\left(-\omega^2-i\frac{2}{\tau_{\alpha}(k)}\omega +\omega_{\alpha}(k)^2\right)D_{\alpha}(k,\omega)
=
\varepsilon a^3 \omega_1^2 E_{0\alpha}(k,\omega),\\
\end{array}
\end{equation}
where the renormalized attenuation rate, 
\begin{equation}
\label{damping111}
\frac{1}{\tau_{\alpha}(k)}=\left\{
\begin{array}{l}
\frac{1}{\tau_0},\;\; for\; 0<kd\pm\omega_1 d/v<2\pi,\\
\frac{1}{\tau_0}+\frac{a^3\omega_1}{2d^3}ImF_{\alpha}(k,\omega_1),\;\; for \;
kd-\omega_1 d/v<0
\;or\; kd+\omega_1 d/a>2\pi,\\
\end{array}
\right.
\end{equation}
and the renormalized self-frequency,
\begin{equation}
\label{freq}
\omega_{\alpha}^2(k)=\omega_1^2\left(1 -\frac{a^3}{d^3}ReF_{\alpha}(k,\omega_1)\right).
\end{equation}

Eq. (\ref{aaa111}) can be easily solved both for inhomogeneous and homogeneous
(when $E_{0\alpha}=0$) case. The general solution of Eq. (\ref{aaa111}) has a form  of sum of the general solution of the homogeneous equation and of a single particular solution of the inhomogeneous equation. The first one includes initial conditions and describes damped self-oscillations with frequency,
\begin{equation}
\label{freq111}
\omega'_{\alpha} =\sqrt{\omega_{\alpha}^2(k)-\frac{1}{\tau_{\alpha}^2(k)}},
\end{equation} 
i.e., for each $k$ and $\alpha$,
\begin{equation}
\label{homogeneous}
D'_{\alpha}(k,t)=A_{\alpha,k}e^{i(\omega'_{\alpha}t+\phi_{\alpha,k})}e^{-t/\tau_{\alpha}(k)},
\end{equation}
where the constants $A_{\alpha, k}$ and $\phi_{\alpha,k}$ are adjusted to the initial conditions. 

   For the inhomogeneous case the particular solution is as follows:
 \begin{equation}
\label{forced}
D''_{\alpha}(k,t)=\varepsilon a^3\omega_1^2E_{0\alpha}(k)e^{i(\gamma t+\eta_{\alpha,k})} \frac{1}{\sqrt{(\omega_{\alpha}^2(k)-\gamma^2)^2+\frac{4\gamma^2}{\tau_{\alpha}^2(k)}}},
\end{equation}
suitably to assumed single Fourier time-component of   $E_{0\alpha}(k,t)=E_{0\alpha}(k)e^{i\gamma t}$, and $tg(\eta_{\alpha,k})=\frac{2\gamma/\tau(\alpha,k)}{\omega(\alpha,k)^2-\gamma^2}$ as usual for a forced oscillator. Let us emphasize that $E_{0\alpha}(k)$ is the real function for  $E_{0\alpha}(ld)^{*}=E_{0\alpha}(ld)=E_{0\alpha}(-ld)$. An appropriate choice of the latter function, in practice a choice of the number of externally excited spheres in the chain, e.g., by suitably focused external excitation, allows for modeling of its Fourier picture $E_{0\alpha}(k)$. This gives the envelope of the wave packet if one inverts  the Fourier transform in the solution given by Eq. (\ref{forced}) back to the position variables.   
For the case of external excitation of only single sphere, the wave packet envelope includes homogeneously all wave vectors $kd\in[0,2\pi)$. The larger number of spheres is simultaneously excited the narrower in $k$ wave packet envelope can be selected. For    $E_{0\alpha}(ld)^{*}=E_{0\alpha}(ld)=E_{0\alpha}(-ld)$ the Fourier transform has the same properties, i.e., $E_{0\alpha}(k)^{*}=E_{0\alpha}(k)=E_{0\alpha}(-k)$ and the latter equality can be rewritten, due to the period  $\frac{2\pi}{d}$ for $k$, as, $ E_{0\alpha}(-k)=E_{0\alpha}(\frac{2\pi}{d}-k)=E_{0\alpha}(k)$.
 The inverse Fourier picture to Eq. (\ref{forced}) (its real part),
\begin{equation}
\label{forcedaaa}
D''_{\alpha} (ld,t)=\int\limits_0^{2\pi/d}dk cos(kld-\gamma t -\eta_{\alpha,k}) \varepsilon a^3\omega_1^2 E_{0\alpha}(k) \frac{1}{\sqrt{(\omega_{\alpha}^2(k)-\gamma^2)^2+\frac{4\gamma^2}{\tau_{\alpha}^2(k)}}}. 
\end{equation}
This integral can be rewritten by virtue of the mean value  theorem in the following form, 
\begin{equation}
\label{forced111}
D''_{\alpha} (ld,t)=\frac{2\pi}{d}  cos(k^{*}ld-\gamma t -\eta_{\alpha,k^{*}}) \varepsilon a^3\omega_1^2 E_{0\alpha}(k^{*}) \frac{1}{\sqrt{(\omega_{\alpha}(k^{*})^2-\gamma^2)^2+\frac{4\gamma^2}{\tau_{\alpha}^2(k^{*})}}}. 
\end{equation}
The above expression describes the undamped wave motion with the frequency $\gamma$ and the velocity, amplitude and phase shift determined by $k^{*}$. The amplitude attains its maximal value at the  resonance, when 
\begin{equation}
\gamma =\omega_{\alpha} (k^{*})\sqrt{1-\frac{2}{(\tau_{\alpha}(k^{*})\omega_{\alpha}(k^{*}))^2}}.
\end{equation}  

In the chain being the subject of persistent time dependent electric field excitation applied to some number (even small number) of spheres, one deals  with undamped wave packed propagating along the whole chain. Such modes depending of particular shaping of the wave packet by specific choice of the chain excitation, may be responsible for experimentally observed long range, practically undamped plasmon-polariton propagation in metallic nano-chains  \cite{kren1999,atwater2003,atwater2003b,atwater2005}. The similar behavior is expected also in ionic micro-chains, due to the shift of all size parameters toward the micro-scale for plasmon-polaritons in ionic systems.  

The self-modes described by Eq. (\ref{homogeneous}) are damped and their propagation depends on appropriately prepared initial conditions admitting nonzero values of $A_{\alpha,k}$. These initial conditions might be prepared by switching off the  time dependent  external electric field exciting initially  some fragment of the chain. The resulting wave packet may embrace the wave-numbers $k$ from some region of $[0,2\pi)$. If only wave-numbers $k$, for which $0<kd\pm \omega_1 d/v< 2\pi$ contribute to the wave packet, its damping is only of Ohmic-type (as is shown in Appendix \ref{a}). The value of corresponding $\frac{1}{\tau_0}$ lowers with growing $a$ (cf. Eq. (\ref{form}))---thus for longer range of these excitations in the chain the larger spheres are favorable. The limiting value of Ohmic losses rate  for large spheres is $\frac{1}{\tau_0}\rightarrow \frac{ v}{2\lambda_B} $  1/s, which gives the maximal range of propagation for irradiation-free  modes of plasmon-polariton. It depends both of mean velocity of ions, $v$ and of their mean free path length, $\lambda_B$; for exemplary $v\sim 1100$ m/s and $\lambda_B\sim 50 \times  10^{-8}$ m, one obtains, for the group velocity of the wave packed assumed as $\sim 0.01c$ m/s,  the range of ionic plasmon-polariton  $\sim 3\times 10^{-3}$ m. The precisely estimated group velocities of plasmon-polaritons   for both polarizations  are presented in Appendix \ref{vg}.

\section{Conclusions}

The presented above analysis, including placed to Appendices quantitative estimations, demonstrates that in chains built of electrolyte finite systems (for the model assumed in form of spheres with radius $a$ in size-scale of micrometers) may be excited collective wave types surface plasmon modes similar to plasmon-polariton in metallic nano-chains. These ionic plasmon-polaritons can efficiently transfer information and energy over relatively large distances and ionic system chains can act as low-losses plasmon wave-guides. These features of the ionic chains, convenient for  plasmon-polariton kinetics, are related to the following properties:
\begin{itemize}
    \item {Ideal cancelation to zero of irradiation losses of collective surface plasmon in the chain for both polarizations of oscillations. This happens due to ideal compensation of irradiation energy losses caused by the Lorentz friction of oscillating charges in each sphere of the chain by radiative income from all neighboring  spheres. The contributions of mutual radiation of all spheres in the chain, including near-, medium- and far-field components of dipole interaction, with precisely accounted for retardation effects, perfectly reduce all irradiation losses and squeeze all the electro-magnetic field to interior of the chain resulting in ideal plasmon-polariton guidance along the ionic chain. The distance of this plasmon-polariton propagation in the chain is limited only by residual Ohmic type losses caused by microscopic scattering of ions, including collisions with other ions, with solvent particles and admixtures and with boundaries of electrolyte spheres. It must be emphasized  that the ionic plasmon damping caused by these irreversible scattering processes are much smaller than the irradiation losses in a single separate electrolyte sphere  with large radius (of several micrometer  order depending on electrolyte parameters). Note that the residual heat losses (Ohmic losses) of plasmon-polaritons in the chain may be compensated by not high outside energetical supporting in the form of persistent excitation or coupling to an active medium (for metal nanochain such an active medium may be the system of quantum dots coupled with metallic nanospheres in the chain and working in regime of spaser \cite{andrianov2012}). The free kinetics of ionic plasmon-polaritons in the chain can reach milimeter scale for the range of propagation, whereas in the case of external compensation of heat dissipation may be practically unlimited.}
\item{The group velocity for ionic plasmon-polaritons may vary in a wide range depending on the electrolyte parameters (mostly on the ion concentration) and on the particularities of  signal wave packet shaping (selecting appropriate envelope for the wave vectors included to the packet of plasmon-polariton modes). Typically, the ionic plasmon-polariton mode velocities fall on the scale of at least  two orders lower than light velocity for ion concentration $\sim 10^{-2} N_0$ ($N_0$ is one-molar concentration) despite the very low  mean value of velocity of ions in electrolytes (due to large their mass). The  value of velocities of ionic plasmon-polaritons in the chains  is correlated also to  its frequency, and  for dense ion concentration is   by $3-4$ orders lower than for plasmon frequency in metals (i.e., $\sim 10^{11-12}$ 1/s in comparison to $10^{15-16}$ 1/s in metals). For small ion concentrations the group velocity can be more significantly  lowered. }  
\end{itemize}

The main properties of ionic plasmon-polaritons as  listed above, are associated also with many specific particularities, as for instance, with a local growth of the group velocities for both plarizations of oscillations in close vicinity of wave vectors satisfying interference condition, $0=kd-\omega  d/v=2\pi$ ($k$ is a wave vector, $d$ is spacing in the chain, $\omega$ is the plasmon-polariton frequency, here $v=c/\sqrt{\varepsilon}$ is the light velocity in electrolyte medium) and occurring due to constructive interference of chain element radiation in far- and medium-field zones. Remarkably, the singular contributions to dipole interaction are effectively quenched by nonlinearity of the dynamics and do not lead to the divergences of the group velocities, though result in local maxima of truncated singularities, sharply cut on the light velocity level (as presented in Appendix \ref{c}). This local enhancement of the group velocity may lead to an occurrence of fast-moving but narrow wave packets of plasmon-polaritons in ionic chains taking advantage of higher velocity of plasmon-polariton modes close to singular points. It must be, however, emphasized that this property is efficiently quenched in finite length chains as for fully development of mentioned singularities the contribution of infinite number (in practice, of large number) of spheres in the chain is needed. The other properties of the plasmon-polariton dynamics in the chain turn out, however, much more robust against shortening of the chain, because the very quick convergences of series representing mutual radiation influence of chain elements, except for singular once (which needs many elements to develop the divergence).

\begin{figure}[h]
\centering
\scalebox{0.35}{\includegraphics{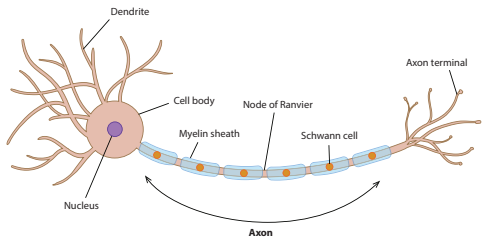}}
\caption{\label{ion-axon} The scheme of the long myelinated axon with chain of Schwann cells of ca 100 $\mu$m length periodically repeated sectors, separated by unmyelinated Ranvier nodes, which gives  of order of 10 000 segments per 1 m of the axon length}  
\end{figure}

In the context of the rich physics of plasmons-polaritons in ionic systems repeating properties of plasmon-polaritons in metallic nano structures, there arises the question on how to verify them experimentally. Ionic systems with sufficiently high ion concentrations are referred rather to liquid electrolytes which do not allow for simple their fragmentation into spheres and chains. Nevertheless, the finite electrolyte systems confined by appropriately formed membranes might serve as the practical realization of model ionic systems. Such matter organization is frequent in biological systems and moreover, in dimension scale which well fits with  size scale typical for ionic plasmon effects, i.e., the  micrometer scale. Cells and internal their components are frequently spheroid-shape structures and ionic effects are of primary significance for their functionality and cell signaling. Looking for some linkage with plasmonic collective effects of ions involved in these systems possess some novelty aspect of approach. The very efficient low-damped and relatively fast kinetics of plasmon-polaritons in ionic system chains is an attractive opportunity especially with by several orders lower frequencies of plasmon oscillations of ions in comparison to high frequency of plasmons in metals, moreover, tunable in wide range by ion concentration. Especially interesting is the question of possibility for modeling of ionic plasmon-polariton signaling in nerve cells, taking into account 
the structure of the long  axons associated with prolate-spheroid shape Schwann cells attaching  the myelin sheath of axon and separated by Ranvier nodes (cf. Fig. \ref{ion-axon}). Though the core of the  axon is a long continuous thin nerve cell insertion, its electric conductivity is relatively poor and there is commonly approved that the  Schwann cell structure (resembling ionic spheroid chain) is crucial in acceleration and maintenance of signals over long distances. The myelin sheath must be sufficiently thick to assure proper functionality of neuron linkage and a myelin deficiency result in slowing down of the signals as it is encountered in Multiple Sclerosis disease. The latter may be caused not by a complete removing of the myelin cover but even only by diminishing its thickness. Note, however, that the of lower thickness myelin sheath still efficiently isolates electrically the core of axon from inter-cell surroundings but probably is insufficient and too thin for dielectric separation needed for plasmon-polariton forming. This may be linked with a distinct role of myelin than only electrical isolation, and its significant function in creation of appropriately  large dielectric/electrolyte periodic segments required to optimize plasmon-polariton kinetics. In possible plasmonic mechanism of signal conducting along the axon it must contribute also   electro-chemical  mechanism of  polarization/depolarization of  those fragments of the membrane  which are unmyelinated in Ranvier nodes. The signal dependent  reversible releasing of ions  trough the lipid membrane sharply enhances and subsequently reduces its polarisation upon the cyclic scheme of neuron activity signal, called as action potential. The incoming to the Ranvier node relatively low signal triggers the opening of $Na^+$ and $K^+$ inter-membrane ion channels, which results in characteristic  larger activation signal due to the transfer through opened gates of ions caused by different their concentrations on both sides of the membrane. The whole cycle takes  the time  of few miliseconds, but initial rising of polarization due to quick $Na^+$ channel happens on single milisecond scale. As the myelin layer surrounded by the Schwann cell prohibits ion across-membrane transfer, the local polarization/depolarization of the internal cytoplasma of axon takes place only in the Ranvier nodes, which strengthen the dipole of the axon fragment wraped by the Schwann cell. If the plasmon-polariton frequency is accomodated to the time scale of activity of ion gate complex in the Ranvier nodes, then the plasmon-polariton mode traversing the axon structure with periodic polarized segments wrapped by Schwann cells can realize the  one-by-one ignition of  Ranvier node sequence. The triggering role of plasmon-polariton  might elucidate thus how the action signal jump between neighboring active nodes in which the dipole oscillation amplitude is raised by functioning of nonlinear block of two ion-channels for $Na^+$ and $K^+$. Due to the nonlinearity (the mutually dependent feedbacks of both ion channels) of this block the polarization amplitude saturates on the 
 constant level (it is kept the same for all nodes).  These active elements of the chain play the role of external energy supply which compensates Ohmic loses of ions and assure long distance transduction of the e-m signal with constant amplitude along the axon.    The velocity of the plasmon-polariton in this chain combined with blocks in Ranvier nodes must be accomodated to plasmon frequency as is illustrated in a simplified manner in Appendix \ref{axon}.   The direct calculus of ion density participating in forming of plasmon-polariton with frequency of sub-milisecond order, may give the velocity of plasmon-polariton of order of 100 m/s (cf. Appendix \ref{axon}), which well fits to observed signal velocity in long myelinated axons. 
 
 The plasmon-polariton  scenario in the ionic chain, though simplified in comparison to the real structure of neurons, might have something in common with the very efficient and energetically economical electric signaling in nerve systems despite rather poor  ordinary conductivity of axons. To avoid slowing down nerve signaling, the plasmon-polariton kinetics in periodic structure of axon is a convenient possibility which  guarantees high level  of its performance: the fast signal propagation without any irradiation losses independently of very low ordinary conductivity, and with practically unlimited range when energy is persistently supplied. The latter is produced by ATP/ADP mechanism in cells, energetically contributing to signal dependent opening and closing of $Na^+$ and $K^+$ channels in Ranvier nodes and next in restoring of steady conditions, i.e., via an active transport of ions across the membrane counter to the ion concentration gradient. Moreover, the coincidence of micrometer scale of axon periodic structure of  Schwann cells (of ca. 100 $\mu$m of length) with typical for ionic chain size-requirements,  supports the suggested plasmon-polariton concept for explanation of action potential transduction along the axon, as presented by the  tentative fitting  in Appendix \ref{axon}. 

It must be also emphasized that plasmon-polariton does not interact with electromagnetic-wave, or equivalently, with photons (even with adjusted energy), which is a consequence of the large difference of group velocity of plasmon-polariton and of photon velocity ($c/\sqrt{\varepsilon}$). The resulting large incommensurability of wave length of photon and plasmon-polariton with the same energy, prohibits a mutual transforming of both excitations owing to the momentum conservation constraints. Therefore plasmon-polariton signaling by dipole oscillations  along the chain, cannot be neither detected  nor perturbed by the external electromagnetic radiation. 
Worth noting is also the  independence of  surface plasmon frequencies  of the temperature in opposition to the volume plasmon frequencies, as it has been shown in the  previous our paper \cite{jacak2014a}. For plasmon-polaritons the temperature influences, however, the mean velocity of ions, $v=\sqrt{\frac{3kT}{m}}$, which enhances Ohmic losses with the temperature rise (cf. Eq. (\ref{form})) and therefore strengthens plasmon-polariton damping. Hence, at higher temperatures  the higher external  energy supplementation is required to maintain the same long range propagation of plasmon-polariton with the same amplitude.    This property is important from point of view of nerve signaling and seems to agree with the observations. 

\vspace{1cm}
{\bf Acknowledgments} Authors 
acknowledge the support of the present work upon the NCN
project\\ no. 2011/03/D/ST3/02643 and  the NCN project no. 2011/02/A/ST3/00116.

\appendix

\section{Calculation of radiative damping of plasmon-polariton in the chain}
\label{a}
Both sides of  Eq.(\ref{tra111111})  can be multiplied by $\frac{e^{i(kld-\omega t)} }{2\pi}$, and next one can perform the summation with respect to sphere positions and the integration over $t$. Taking  into account that, 
\begin{equation}
\begin{array}{l}
\frac{1}{2\pi}\int\limits_\infty^\infty \sum\limits_{l=-N}^{N}
D_{\alpha}\left(\pm md+ld;t-\frac{md}{v}\right) e^{-i(kld-\omega t)}\\
=e^{i\left(\mp kmd +\omega\frac{md}{v}\right)}D_{\alpha}(k,\omega),\\
\end{array}
\end{equation}
one obtains thus the following equation in Fourier representation (the discrete Fourier transform  for sphere positions and the continuous Fourier transforms for time),
\begin{equation}
\label{a1}
\begin{array}{l}
\left(-\omega^2-i\frac{2}{\tau_0}\omega +\omega^2_1\right)D_{\alpha}(k,\omega)\\
=\omega_1^2\frac{a^3}{d^3}F_{\alpha}(k,\omega)D_{\alpha}(k,\omega)+
\varepsilon a^3 \omega_1^2 E_{0\alpha}(k,\omega),\\
\end{array}
\end{equation}
where $k=\frac{2\pi n}{2Nd}, \;\; n=0,1,...,2N$,   i.e., $kd\in [0,2\pi)$   due to periodicity of the chain with equidistant separation $d$ of spheres (separation between sphere centers), and the form of $k$ is due to  Born-Karman boundary condition with the period $L=2Nd$. For $N \rightarrow \infty $ (infinite chain limit) $k$ is a quasi-continuous variable. In  Eq. (\ref{a1}), 
\begin{equation}
\label{podluzneipoprzeczne}
\begin{array}{l}
F_z(k,\omega)=4\sum\limits_{m=1}^\infty \left(\frac{cos(mkd)}{m^3}cos(m\omega d/v)+\omega d /v \frac{cos(mkd)}{m^2}sin(m\omega d/v)\right)\\
+2i \left[\frac{1}{3}(\omega d /v)^3+2\sum\limits_{m=1}^\infty \left(\frac{cos(mkd)}{m^3}sin(m\omega d/v)\right.\right.\\
\left.\left. -\omega d/v\frac{cos(mkd)}{m^2}cos(m\omega d/v)\right)\right],\\
F_{x(y)}(k,\omega)=-2\sum\limits_{m=1}^\infty \left(\frac{cos(mkd)}{m^3}cos(m\omega d/v)+\omega d /v \frac{cos(mkd)}{m^2}sin(m\omega d/v)\right.\\
\left. -(\omega d/v)^2\frac{cos(mkd)}{m}cos(m\omega d/v)\right)\\
-i \left[-\frac{2}{3}(\omega d /v)^3+2\sum\limits_{m=1}^\infty \left(\frac{cos(mkd)}{m^3}sin(m\omega d/v)\omega d/v\frac{cos(mkd)}{m^2}cos(m\omega d/v)\right.\right.\\
\left.\left.-(\omega d/v)^2\frac{cos(mkd)}{m}sin(m\omega d/v)\right)\right].\\
\end{array}
\end{equation}

Some  summations in the above equations  can be performed analytically \cite{gradst},
\begin{equation}
\label{wzory}
\left\{ \begin{array}{l}
\sum\limits_{m=1}^\infty \frac{sin(mz)}{m}=\frac{\pi -z}{2},\;\; for\;\;0<z<2\pi,\\
 \sum\limits_{m=1}^\infty \frac{cos(mz)}{m}=\frac{1}{2}ln\left(\frac{1}{2-2 cos(z)}\right),\\
\sum\limits_{m=1}^\infty \frac{cos(mz)}{m^2}=\frac{\pi^2}{6} -\frac{\pi}{2}z
+\frac{1}{4}z^2,\;\; for\;\;0<z<2\pi,\\
\sum\limits_{m=1}^\infty \frac{sin(mz)}{m^3}=\frac{\pi^2}{6}z -\frac{\pi}{4}z^2
+\frac{1}{12}z^3,\;\; for\;\;0<z<2\pi.\\
\end{array}
\right.
\end{equation}
Using the above formulae one can show that if $0<kd\pm\omega d/v<2\pi$,   then,
\begin{equation}
\label{podluzne}
\begin{array}{l}
ImF_z(k,\omega)=2\sum\limits_{m=1}^\infty \left[\frac{sin(m(kd+\omega d/v))-sin(m(kd-\omega d/v))}{m^3} \right.\\
\left.-(\omega d/v)\frac{cos(m(kd+\omega d/v))+cos(m(kd-\omega d/v))}{m^2}\right]+\frac{2}{3}(\omega d/v)^3\\
=2\left[\frac{\pi^2}{6}(kd+\omega d/v)-\frac{\pi}{4}(kd+\omega d/v)^2+\frac{1}{12}(kd +\omega d /v)^3\right]\\
-2\left[\frac{\pi^2}{6}(kd-\omega d/v)-\frac{\pi}{4}(kd-\omega d/v)^2+\frac{1}{12}(kd -\omega d /v)^3\right]\\
-2(\omega d/v)\left[\frac{\pi^2}{6}-\frac{\pi}{2}(kd+\omega d/v)+\frac{1}{4}(kd+\omega d/v)^2\right]\\
-2(\omega d/v)\left[\frac{\pi^2}{6}-\frac{\pi}{2}(kd-\omega d/v)+\frac{1}{4}(kd-\omega d/v)^2\right]+\frac{2}{3} (\omega d/v)^3\equiv 0.\\
\end{array}
\end{equation}

\begin{figure}[h]
\centering
\scalebox{0.30}{\includegraphics{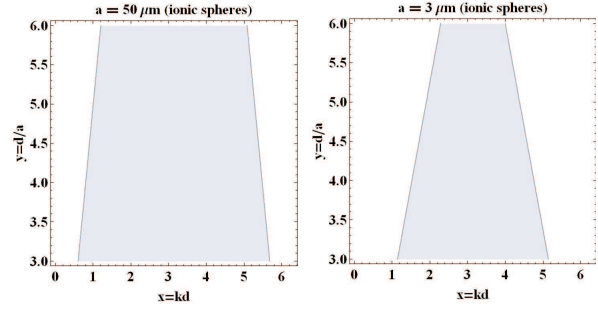}}
\caption{\label{figk2} The region (gray, $0<kd\pm \omega_1 d/v<2\pi$) in where the radiation losses vanish for infinite chains of electrolyte spheres with various sphere radii and chain-separation $d/a\in[3,6]$, for $\omega=\omega_1=1.2\times 10^{12}\; (3.8\times 10^{13})$ 1/s for ion concentration $n=10^{-3(-2)}  N_0$, $N_0$ is one-molar concentration---left(right), $v=c$}
\end{figure}

\begin{figure}[h]
\centering
\scalebox{0.35}{\includegraphics{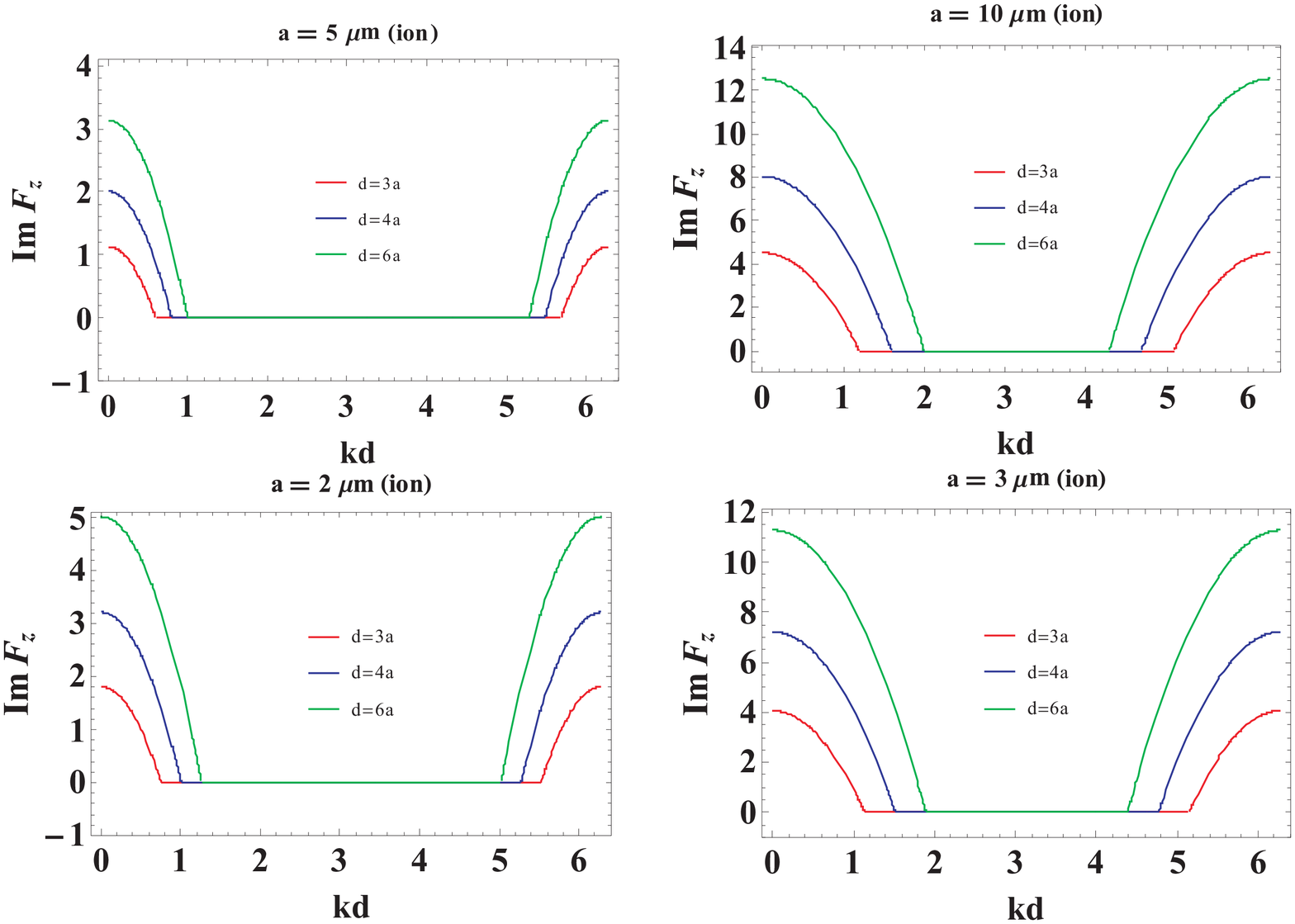}}
\caption{\label{figk1} The function $ImF_{z}(k;\omega=\omega_1)$ for infinite chains of ionic spheres with radius $a$ and chain-separations $d=3a,\;4a,\;5a$
for effective ions mass $m=10^4 m_e$, effective charge $q=3e$, dielectric susceptibility of surroundings  $\varepsilon=2$, $T=300$ K, ion concentration  $n=10^{-3}N_0$ (upper), $n=10^{-2}N_0$ (lower), $N_0$ is one-molar electrolyte concentration}
\end{figure}

However, if $kd-\omega d/v <0$  or $kd+\omega d/v>2\pi $  for some values of the wave vector $k$, the more general  formula must be used  (by application of  Heaviside step function one can extend the formulae (\ref{wzory}) to the second period of their left sides). This extended form for $ImF_z(k,\omega)$ is as follows (here we use dimensionless variables $x=kd,\;y=d/a$),
\begin{equation}
\label{podluzne111}
\begin{array}{l}
ImF_z(k,\omega)=\Theta(2\pi-x-\omega ya/v)2\left[\frac{\pi^2}{6}(x+\omega ya /v)-\frac{\pi}{4}(x+\omega ya/v)^2+\frac{1}{12}(x+\omega ya/v)^3\right]\\
\Theta(-2\pi+x+\omega ya/v)2\left[\frac{\pi^2}{6}(x+\omega ya /v-2\pi)-\frac{\pi}{4}(x+\omega ya/v-2\pi)^2+\frac{1}{12}(x+\omega  ya/v-2\pi)^3\right]\\
-\Theta(x-\omega ya/v)2\left[\frac{\pi^2}{6}(x-\omega ya /v)-\frac{\pi}{4}(x-\omega ya/v)^2+\frac{1}{12}(x-\omega ya/v)^3\right]\\
-\Theta(-x+\omega ya/v)2\left[\frac{\pi^2}{6}(x-\omega ya /v+2\pi)-\frac{\pi}{4}(x-\omega ya/v+2\pi)^2+\frac{1}{12}(x-\omega  ya/v+2\pi)^3\right]\\
-\Theta(2\pi-x-\omega ya/v)2 (\omega a y/v)\left[\frac{\pi^2}{6}-\frac{\pi}{2}(x+\omega ya/v)+\frac{1}{4}(x+\omega  ya/v)^2\right]\\
-\Theta(-2\pi+x+\omega ay/v)2(\omega ay/v) \left[\frac{\pi^2}{6}-\frac{\pi}{2}(x+\omega ya/v-2\pi)+\frac{1}{4}(x+\omega  ya/v-2\pi)^2\right]\\
-\Theta(-x+\omega ay/v)2(\omega ay/v) \left[\frac{\pi^2}{6}-\frac{\pi}{2}(x-\omega ya/v+2\pi)+\frac{1}{4}(x-\omega  ya/v+2\pi)^2\right]\\
-\Theta(x-\omega ay/v)2(\omega ay/v) \left[\frac{\pi^2}{6}-\frac{\pi}{2}(x-\omega ya/v)+\frac{1}{4}(x-\omega  ya/v)^2\right]+\frac{2}{3}(\omega a y/v)^3.\\
\end{array}
\end{equation}
The  function given by Eq. (\ref{podluzne111}) is depicted in  Fig. \ref{figk1}. The expression (\ref{podluzne111}) allows to account for the inconsistence  of periodic functions given by the sums of sines and cosines with non periodic Lorentz friction term and  disagreement  of  arguments $kd\pm \omega d/v$ of trigonometric functions out of the first period.  In Fig \ref{figk2} we have plotted the  solution of the equation $(kd-\omega d/v)(kd+\omega d/v -2\pi)=0$,which determines the region for $kd$ (denoted by $x$) versus $d/a$ (denoted by $y$) inside which  the exact cancellation of the Lorentz friction by radiative energy income from other nanospheres takes place. In Fig. \ref{figk1}, the comparison of this cancellation for various nanosphere diameters is presented, for longitudinal polarization of plasmon collective excitations. 

The similar analysis can be done for transversal polarization, i.e., for $Im F_{x(y)}(k,\omega)$. This function is exactly zero only in the region for arguments $0<kd-\omega d/v<2\pi$ and $0<kd+\omega d/v<2\pi$, where one can write,
\begin{equation}
\label{poprzeczne}
\begin{array}{l}
ImF_{x(y)}(k,\omega)=-\sum\limits_{m=1}^\infty \left[\frac{sin(m(kd+\omega d/v))-sin(m(kd-\omega d/v))}{m^3} \right.\\
\left.-(\omega d/v)\frac{cos(m(kd+\omega d/v))+cos(m(kd-\omega d/v))}{m^2}\right.\\
\left.-(\omega d/v)^2
\frac{sin(m(kd+\omega d/v))-sin(m(kd-\omega d/v))}{m}\right]
+\frac{2}{3}(\omega d/v)^3\\
=-\left[\frac{\pi^2}{6}(kd+\omega d/v)-\frac{\pi}{4}(kd+\omega d/v)^2+\frac{1}{12}(kd +\omega d /v)^3 \right]   \\
+\left[\frac{\pi^2}{6}(kd-\omega d/v)-\frac{\pi}{4}(kd-\omega d/v)^2+\frac{1}{12}(kd -\omega d /v)^3\right]\\
+(\omega d/v)\left[\frac{\pi^2}{6}-\frac{\pi}{2}(kd+\omega d/v)+\frac{1}{4}(kd+\omega d/v)^2\right]\\
+(\omega d/v)\left[\frac{\pi^2}{6}-\frac{\pi}{2}(kd-\omega d/v)+\frac{1}{4}(kd-\omega d/v)^2\right]\\
\frac{1}{2}(\omega d/v)^2[\pi -kd -\omega d/v]-\frac{1}{2}[\pi -kd+\omega d/v]+\frac{2}{3} (\omega d/v)^3\equiv 0.\\
\end{array}
\end{equation}

\begin{figure}[h]
\centering
\scalebox{0.35}{\includegraphics{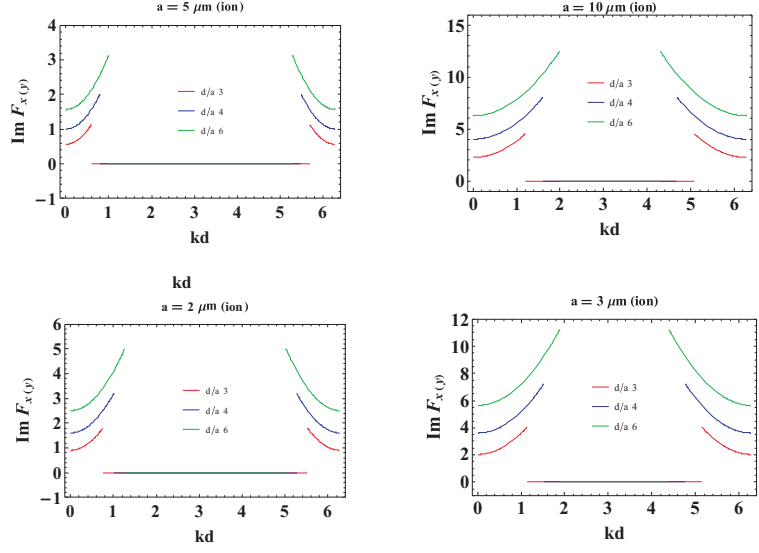}}
\caption{\label{figk3} The function $ImF_{x(y)}(k;\omega=\omega_1)$
for infinite chains of ionic spheres with radius $a$ and chain-separations $d=3a,\;4a,\;5a$
for effective ions mass $m=10^4 m_e$, effective charge $q=3e$, dielectric susceptibility of surroundings  $\varepsilon=2$, $T=300$ K, ion concentration  $n=10^{-3}N_0$ (upper), $n=10^{-2}N_0$ (lower), $N_0$ is one-molar electrolyte concentration}
\end{figure}

Nevertheless, outside the region $0<kd\pm \omega d/v<2\pi$  the value of $Im F_{x(y)}$  is not zero, as it is demonstrated in Fig. \ref{figk3}, and can be accounted for by the formula ($x=kd$, $y=d/a$),
\begin{equation}
\label{poprzeczne111}
\begin{array}{l}
ImF_{x(y)}(k,\omega)=-\Theta(2\pi-x-\omega ya/v)\left[\frac{\pi^2}{6}(x+\omega ya /v)-\frac{\pi}{4}(x+\omega ya/v)^2+\frac{1}{12}(x+\omega ya/v)^3\right]\\
-\Theta(-2\pi+x+\omega ya/v)\left[\frac{\pi^2}{6}(x+\omega ya /v-2\pi)-\frac{\pi}{4}(x+\omega ya/v-2\pi)^2+\frac{1}{12}(x+\omega  ya/v-2\pi)^3\right]\\
+\Theta(x-\omega ya/v)\left[\frac{\pi^2}{6}(x-\omega ya /v)-\frac{\pi}{4}(x-\omega ya/v)^2+\frac{1}{12}(x-\omega ya/v)^3\right]\\
+\Theta(-x+\omega ya/v)\left[\frac{\pi^2}{6}(x-\omega ya /v+2\pi)-\frac{\pi}{4}(x-\omega ya/v+2\pi)^2+\frac{1}{12}(x-\omega  ya/v+2\pi)^3\right]\\
+\Theta(2\pi-x-\omega ya/v)(\omega a y/v)\left[\frac{\pi^2}{6}-\frac{\pi}{2}(x+\omega ya/v)+\frac{1}{4}(x+\omega  ya/v)^2\right]\\
+\Theta(x-\omega ay/v)(\omega ay/v) \left[\frac{\pi^2}{6}-\frac{\pi}{2}(x-\omega ya/v)+\frac{1}{4}(x-\omega  ya/v)^2\right]\\
+\Theta(-2\pi+x+\omega ay/v)(\omega ay/v) \left[\frac{\pi^2}{6}-\frac{\pi}{2}(x+\omega ya/v-2\pi)+\frac{1}{4}(x+\omega  ya/v-2\pi)^2\right]\\
+\Theta(-x+\omega ay/v)(\omega ay/v) \left[\frac{\pi^2}{6}-\frac{\pi}{2}(x-\omega ya/v+2\pi)+\frac{1}{4}(x-\omega  ya/v+2\pi)^2\right]\\
+\Theta(2\pi -x - \omega ya/v)\frac{1}{2}(\omega ay/v)^2[\pi-x-\omega y a/v]
+\Theta(-2\pi +x +\omega y a/v)\frac{1}{2}(\omega a y/v)^2[3\pi -x-\omega y a/v]\\
-\Theta(x-\omega ya/v)\frac{1}{2}(\omega a y/v)^2[\pi -x+\omega y a/v]
-\Theta(-x+\omega a y/v)^2[-\pi -x+\omega y a/v]
+\frac{2}{3}(\omega a y/v)^3.\\
\end{array}
\end{equation}

\begin{figure}[h]
\centering
\scalebox{0.22}{\includegraphics{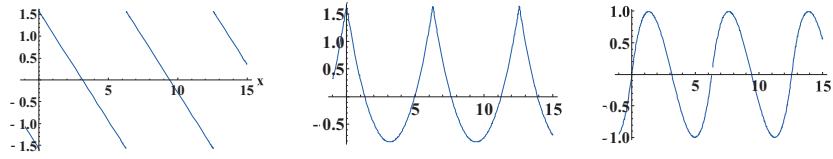}}
\caption{\label{figk4} Sums $\sum\limits_{n=1}^{\infty}\frac{sin(nx)}{n}$ (left), $\sum\limits_{n=1}^{\infty}\frac{cos(nx)}{n^2}$ (center), $\sum\limits_{n=1}^{\infty}\frac{sin(nx)}{n^3}$ (right), for $x\in(-1,15)$ }
\end{figure}

This function is plotted in Fig. \ref{figk3}. The discontinuity jump on the border between the regions with vanishing radiative damping and with nonzero radiative attenuation is caused by discontinuous function $\sum\limits_{n=1}^\infty \frac{sin(nz)}{n}$ (cf. Fig. \ref{figk4}) entering $ImF_{x(y)}$ but not $ImF_z$, cf. Eq.(\ref{podluzneipoprzeczne}).

\begin{figure}[h]
\centering
\scalebox{0.35}{\includegraphics{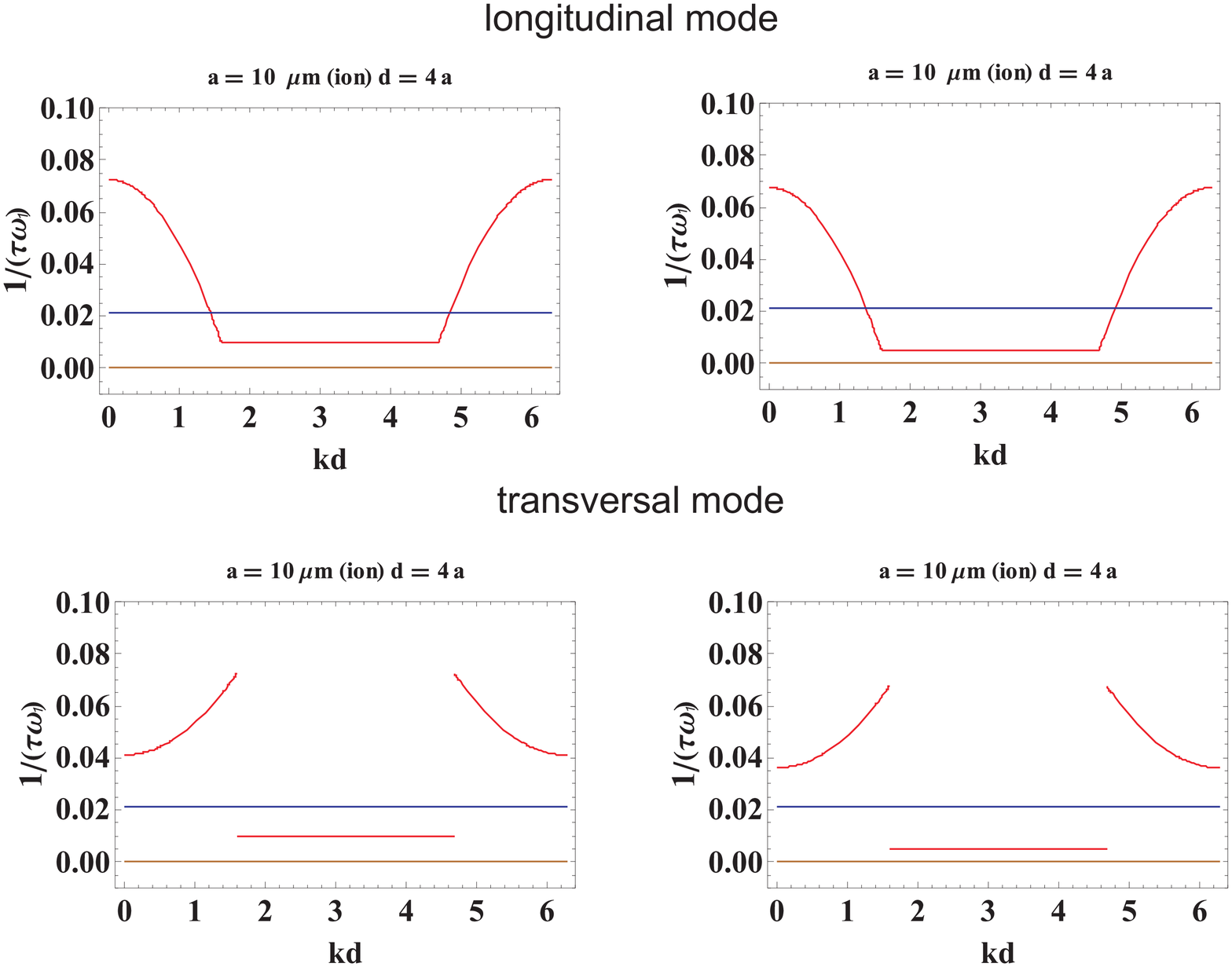}}
\caption{\label{figk91} Comparison of various contributions to  collective plasmon damping in the chain---red line corresponds to total damping ratio (in dimensionless units) $\frac{1}{\omega_1 \tau_{z}}=
\frac{1}{\omega_1 \tau_0} +\frac{a^3}{2d^3}ImF_z(k)$, the bottom of this line plot displays only Ohmic contribution at that segment of the $k$ period where all e-m losses vanish, the blue line indicates the Lorentz friction value (in the same dimensionless units)  $\frac{1}{3}\left(\frac{\omega_1 a}{v}\right)^3$;
for effective ions mass $m=10^4 m_e$, effective charge $q=3e$, dielectric susceptibility of surroundings  $\varepsilon=2$, $T=300$ K, ion concentration, $n=10^{-3}N_0$, $N_0$ is one-molar electrolyte concentration and ion mean free path $\lambda_B=5\times 10^{-9}$ m (left) $\lambda_B= 10^{-8}$ m (right) }
\end{figure}

In order to compare  the magnitudes of various contributions to the damping of collective plasmons in the chain  let us plot dimensionless values, for the longitudinal polarization, $\frac{1}{\omega_1 \tau}=
\frac{1}{\omega_1 \tau_0} +\frac{a^3}{2d^3}ImF_z(k)$ (in red in Fig. \ref{figk91}) in comparison to the Lorentz friction contribution $\frac{1}{3}\left(\frac{\omega_1 a}{v}\right)^3$ (blue line in Fig \ref{figk91}, cf. Eq. (\ref{damping111})). In this figure one can note that for exemplary spheres the Lorentz term is much larger  than the Ohmic attenuation (the bottom of red line, cf. also Eq. (\ref{form})). For large spheres the Ohmic damping is governed by the ratio  $\frac{v}{\lambda_B}$ in Eq. (\ref{form}), which is small due to extremely small mean velocity of carriers in the electrolyte despite also small mean free path. The same can be demonstrated for the transversal polarization, as it is shown in Fig. \ref{figk91}.

\section{Calculation of self-frequencies and group velocities of plasmon-polaritons in ionic electrolyte chain}

\label{vg}

According to Eqs (\ref{freq})
and (\ref{freq111}) for self-frequencies of plasmon-polariton in the chain one can write out their explicit forms using expressions for $F_{\alpha}(k,\omega_1)$ given by Eq (\ref{aaapodluzneipoprzeczne}). The real part of the functions $F_{\alpha}(k,\omega_1)$ renormalizes the corresponding frequencies for both polarization and they attain the following forms (according to Eq (\ref{freq})):
\begin{equation}
\label{czestoscpodluzna}
\begin{array}{l}
\omega_z^2 (k)=\omega_1^2\left[1-\frac{a^3}{d^3} 
4\sum\limits_{m=1}^\infty \left(\frac{cos(mkd)}{m^3}cos(m\omega_1 d/v)+\omega_1 d /v \frac{cos(mkd)}{m^2}sin(m\omega_1 d/v)\right)\right]\\
=\omega_1^2\left[1-\frac{a^3}{d^3} 
2\sum\limits_{m=1}^\infty \left(\frac{cos(m(x+\omega_1ya/v))+cos(m(x-\omega_1ya/v))}{m^3}+
\omega_1 ya /v \frac{sin(m(x+\omega_1ya/v))-sin(m(x-\omega_1ya/v))}{m^2}\right)\right],\\
\end{array}
\end{equation}
\begin{equation}
\label{czestoscpoprzeczna}
\begin{array}{ll}
\omega_{x(y)}^2 (k)&=\omega_1^2\left[1+\frac{a^3}{d^3}
2\sum\limits_{m=1}^\infty \left(\frac{cos(mkd)}{m^3}cos(m\omega_1 d/v)+\omega_1 d /v \frac{cos(mkd)}{m^2}sin(m\omega_1 d/v)\right.\right.\\
&\left.\left. -(\omega_1 d/v)^2\frac{cos(mkd)}{m}cos(m\omega_1 d/v)\right)
 \right]\\
&=\omega_1^2\left[1+\frac{a^3}{d^3}
\sum\limits_{m=1}^\infty \left(\frac{cos(m(x+\omega_1ya/v)+cos(m(x-\omega_1ya/v))}{m^3}
+\omega_1ay/v \frac{sin(m(x+\omega_1 ya/v)) -sin(m(x-\omega_1 ya/v))}{m^2}\right.\right.\\
&\left.\left. -(\omega_1 ay/v)^2\frac{cos(m(x+\omega_1ya/v)) +cos(m(x-\omega_1 ya/v))}{m}\right)
 \right],\\
\end{array}
\end{equation}
where $x=kd$ and $y=d/a$.

\begin{figure}[h]
\centering
\scalebox{0.35}{\includegraphics{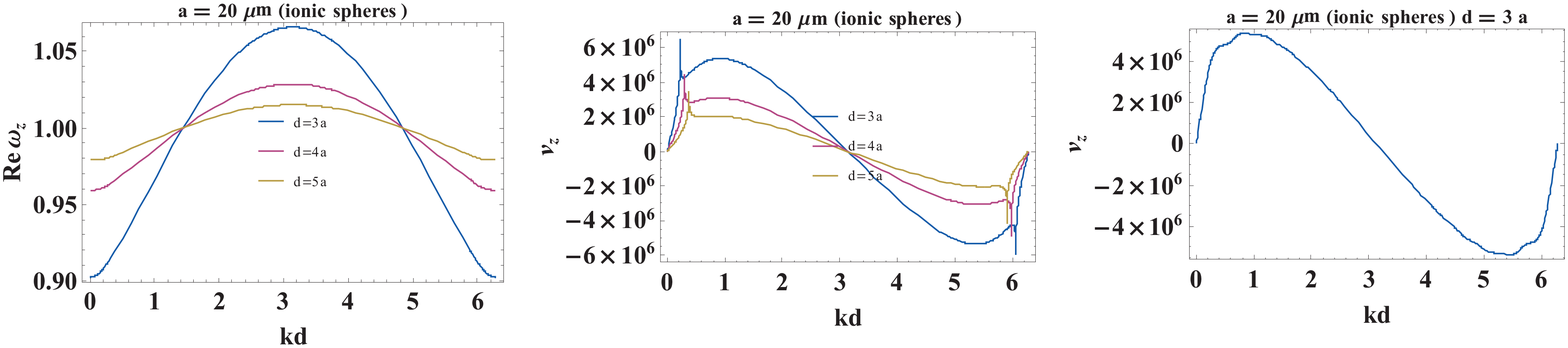}}
\caption{\label{fig6k} The self-frequency $\omega'_{z}(k)$  (in $\omega_1$ units) and the group velocity (in m/s) of plasmon polariton with longitudinal polarization  in infinite chain of electrolyte spheres with radius $a=20$ $\mu$m  and chain-separation $d=3, \;4a,\;5a$, ion concentration $n=10^{-3}N_0$, $T=300$ K, $\omega_1=1.2\times 10^{12}$ 1/s; in the right panel effective removal  of logarithmic singularity in the group velocity due to  the far-field  contribution in the finite chain of 15 spheres;  the logarithmic singularity for the infinite chain  in the perturbation formula for $v_z$ (though not in formula for $\omega'_z$) leads to local exceeding of $c$---this artifact of perturbation approach is removed by  the exact solution, cf. Appendix \ref{c}} 
\end{figure}

\begin{figure}[h]
\centering
\scalebox{0.32}{\includegraphics{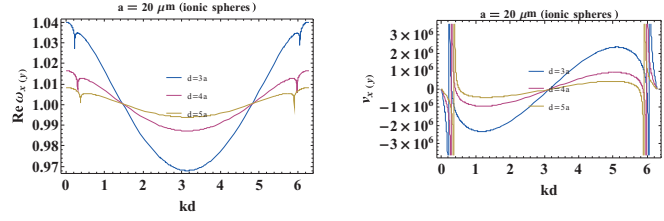}}
\caption{\label{fig5k} The self-frequency $\omega'_{x(y)}(k)$  (in $\omega_1 $ units) and the group velocity (in m/s) of plasmon polariton  with transversal  polarization  in infinite chains of  electrolyte spheres with radius $a=20$ $\mu$m  and chain-separation $d=3a,\;4a,\;5a$, ion concentration $n=10^{-3}N_0$, $T=300$ K, $\omega_1=1.2\times 10^{12}$ 1/s;
the logarithmic singularity in  $\omega'_{x(y)}(k)$  due to constructive interference of far-field contribution of dipole interaction, and the related hyperbolic singularity in transversal group velocity are, however, an  artifact of perturbation method of solution of Eq. (\ref{aaa1})---in the exact solution of this equation the singularities are quenched to  small local  minima only, cf. Appendix \ref{c}}
\end{figure}

The shift of self-frequencies of plasmon-polaritons cased by their attenuation is accounted for by the formula (\ref{freq111}) where $\omega_{\alpha} ( k)$ is given by Eqs (\ref{czestoscpodluzna}) and (\ref{czestoscpoprzeczna})  where $\tau_{\alpha} (k)$ has a form as in Eq. (\ref{damping111}). Taking into account the explicit form for $ImF_{\alpha}(k,\omega_1)$, i.e., the expressions (\ref{podluzne111}) and (\ref{poprzeczne111}), one can easily calculate the self-frequencies $\omega'_{\alpha}(k)$ -- these functions are shown in  Figs \ref{fig6k} and \ref{fig5k} for longitudinal and transversal polarizations, respectively.

Note that for the transversal polarization in Eq. (\ref{czestoscpoprzeczna})
the sum,
\begin{equation}
\label{b3}
\sum\limits_{m=1}^\infty\frac{cos(m(x+\omega_1ya/v)) +cos(m(x-\omega_1 ya/v))}{m}, 
\end{equation}
can be performed analytically according to the formula (\ref{wzory}) resulting in the contribution,
\begin{equation}
\label{b4}
-\frac{1}{2}ln[ (2-2cos(x+\omega_1ya/v))(2-2cos(x-\omega_1ya/v))]
\end{equation}
 (the other sums in Eqs (\ref{czestoscpodluzna}) and (\ref{czestoscpoprzeczna}) have to be done numerically). This logarithmic singularity in self-frequencies for transversal plasmon-polaritons on the rim of the region $0<x-\omega_1ya/v<2\pi$ (inside which radiative damping vanishes) is indicated in Fig. \ref{fig5k} left. This singularity causes hyperbolic discontinuity in transversal group velocity (cf. Fig. \ref{fig5k} right).   Nevertheless, the logarithmic singularity in self-energy and the related hyperbolic discontinuity of transversal braid group turn out to be an artifact of perturbation solution of Eq. (\ref{aaa1}). The exact numerical solution of this equation demonstrates an effective quenching of the logarithmic singularity to small local minimum resulting in finite group velocity discontinuity, as it is shown in  Appendix \ref{c}. This property of transversal propagation of plasmon-polaritons in the chain caused by constructive interference of the far-field component of dipole interactions between spheres was analyzed for metallic chain  numerically in \cite{markel2007} and commented  in \cite{jacak2013,jacak2014}.
The numerical studies of plasmon-polariton propagation in metallic nano-chain \cite{markel2007} indicated very narrow and weak but long range mode besides the wide spectrum of quickly damped modes. This 'fainting' long range mode can be associated with  interference of far-field part of dipole-dipole interaction of nanospheres in the chain, resulting in local enhancement of transversal group velocity in narrow vicinity of singularity points. The same can be addressed to electrolyte ionic  chains.

In order to find the group velocities of particular self-modes of plasmon-polariton in the chain, the derivative of $\omega'_{\alpha}( k)$ with respect to $k$ must be performed, which according to the expressions (\ref{czestoscpodluzna}), (\ref{czestoscpoprzeczna}), (\ref{freq111}) and (\ref{damping111}) is a straightforward though an extended calculus. The sums in formulae for $\omega_{\alpha}(k)$ still cannot be done analytically except for the sum with linear denominator in (\ref{czestoscpoprzeczna}). The resulting group velocity calculated numerically for both polarizations and for $kd\in[0,2\pi)$ and $d/a\in[3,10]$ are presented in Figs \ref{fig6k} and \ref{fig5k} (for exemplary chain parameters and  for both polarizations).

\section{Exact solution of Eq. (\ref{aaa1})---the problem of the logarithmic divergence of the far-field contribution to self-frequency for transversal polarization of plasmon polariton and of the medium-field contribution to  group velocities for both polarizations}
\label{c}

The solution of  Eq. (\ref{aaa1}) in the homogeneous case gives  the Fourier exponent $\omega_{\alpha}(k)$, which  is  a complex function, in general. Its  imaginary part defines plasmon-polariton damping rate, whereas  the real part of  the $\omega$ function  defines self-frequency of particular $k$ mode with $\alpha$ polarization.  One can next find the group velocities for particular modes taking   the derivative of the real part of $\omega_{\alpha}(k)$   with respect to the wave vector $k$. As we have noticed above, the far-field contribution to the dipole interaction in the chain produces a logarithmically singular term in Eq. (\ref{aaa1}) of the form given by Eqs (\ref{b3})-(\ref{b4}).
Because of infinite divergence of this term
one can not apply the perturbation method for solution Eq. (\ref{aaa1}), in the vicinity  to the singular points in $k$ domain. 
The perturbative method of solution of Eq. (\ref{aaa1}) transfers this logarithmic singularity onto solution for self-frequency in the singular point $k$. The derivative of this frequency, i.e., the group velocity of the singular mode $k$, acquires in this way  the hyperbolic singularity. This happens for perturbative formula for the transversal group velocity. These infinite divergences are visible in Fig.\ref{fig5k} (left) close to the edges of the domain $kd\in[0, 2\pi)$. 

For the longitudinal polarization upon perturbation approach one encounters also infinite singularity of the group velocity in the same $k$ as for the transversal polarization, though without any singularity in the longitudinal self-frequency. This astonishing singularity in the group velocity is of the  logarithmic type and arises actually for both polarizations, when  one takes the derivative with respect to $k$ from Eqs (\ref{czestoscpodluzna})-(\ref{czestoscpoprzeczna}). The contribution to self-energies for both polarizations sourced by medium-field irradiation term 
with the factor $\sum \frac{cos(nx)}{n^2}$,  after taking the derivative with respect to $x$ ($x=kd$) is divergent. For the longitudinal polarization this divergence is visible in the Fig. \ref{fig6k}. For the transversal polarization this logarithmic singularity interferences with above described hyperbolic one.  These all singularities occur at isolated points for which $kd \pm \omega_1d/v=l\pi$ ($l$ - integer). These singularities are apparent artifact of the perturbation approach because lead to local exceeding of light velocity. 

In order to resolve and clarify this  problem of unphysical divergence one has to solve  Eq. (\ref{aaa1}) exactly not applying the perturbation technique. This is clear, because of divergence of the expression ({\ref{b3}) the corresponding contributions cannot be treated  as a small perturbation. The exact solution of Eq. (\ref{aaa1}) can be found numerically. The result of numerical solution (by the Newton type algorithm) for self-frequencies with  both polarization and for whole $k$ domain  is plotted in Fig. \ref{55}.
\begin{figure}[h]
\centering
\scalebox{0.35}{\includegraphics{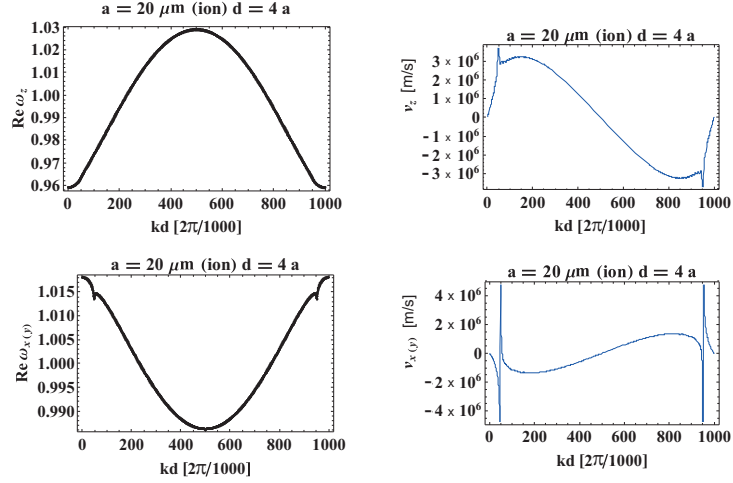}}
\caption{\label{55} The exact solutions for self-frequencies  for  the longitudinal and  transversal modes of plasmon-polariton in the ionic chain ($\omega$ in $\omega_1$ units) obtained by   the exact solution of Eq. (\ref{aaa1}) in 1000 points on the sector $kd\in [0,2\pi$)---left; and corresponding group velocities for both polarizations---right}  
\end{figure}

By the comparison with the  corresponding plots obtained within the perturbation method and those in   Fig \ref{55},  we notice  that   the exact solutions for self-frequencies  do not differ significantly  from those obtained in the perturbation manner. Nevertheless,  the exact solutions for the group velocities do not have any singularities. As it follows from the thorough analysis \cite{jacak2014} in the exact solutions  for group velocities all divergences are cut at the level of light velocity. 
This quenching of singularities in the way protecting not exceeding the light velocity is the manifestation of the Lorentz relativistic invariance of plasmon-polariton dynamics description. 
 The retardation of electric signals prohibits the collective excitation group velocity to exceed the light velocity.  This quenching concerns  infinite singularities which occur in perturbation expressions for self-energy and next in perturbation formulae for group velocities.  The exact solution of the Lorentz invariant dynamical equation  inherently posses also this property.  Exact self-energies have suitably regularized their dependence  with respect to $k$, that their derivatives  do not exceed $v=\frac{c}{\sqrt{\varepsilon}}$.
It is worth emphasizing for the sake of completeness of the description, that inclusion  of magnetic field of dipoles does not modify this scenario,  because the magnetic field contribution to self-energies is a few orders lower in comparison to the electric  field contribution due to the  velocity of ions being also similarly  lower in comparison to light velocity, which significantly reduces the Lorentz force caused by the magnetic field. Therefore, the magnetic field of the dipoles \cite{lan,jackson1998}, 
\begin{equation}
B_{\omega}=ik(\mathbf{D}_{\omega} \times \mathbf{n})\left( \frac{ik}{r_0} - \frac{1}{r_0^2}\right)e^{ik r_0},
\end{equation}
though contributing to the far-field and medium-field parts of plasmon-polariton  self-energies, does not change significantly the appropriate terms caused by the electric field and causes only  corrections  practically negligible. 

\section{Fitting of plasmon-polariton features  to parameters of an axon in the peripheral nervous system}

\label{axon}

The bulk plasmon frequency is linked to ion parameters via the formula,
$\omega_p=\sqrt{\frac{q^2 n}{\varepsilon_0 m}}$, (in Gauss units, $\omega_p=\sqrt{\frac{q^2 n 4 \pi}{m}}$), 
where $q$ is the charge of the ion, $n$ is the ion concentration, $\varepsilon_0$ is the dielectric susceptibility of the vacuum, $m$ is the mass of the ion. For model we assume the charge of ions,  $q=1.6 \times 10^{-19}$ C (the electron charge), and their mass, $m=10^4  m_e$, where $m_e=9.1 \times 10^{-31}$ kg is the mass of the electron, $n=2.1 \times 10^{14}$ 1/m$^3$, and thus we obtain  the Mie frequency for ionic dipole oscillations,  $\omega_1
\simeq 0.1 \frac{\omega_p}{\sqrt{3 \varepsilon}}\simeq 4 \times 10^6$ 1/s, where for the MHz frequency, the  relative permittivity of  water $\varepsilon \simeq 80$ \cite{epsilon} (though for the visible light frequencies it drops, staring  the fall  at ca 10 GHz, to the final value ca. 1.7,  corresponding to the refractive index for water $\eta\simeq \sqrt{\varepsilon_1}=1.33$). For the strongly prolate and thin  ionic inner cord we reduced the longitudinal Mie frequency by factor 0.1 \cite{elipsa1,elipsa2} in comparison to the isotropic spherical case, when it was $\omega_1=\frac{\omega_{p}}{\sqrt{3 \varepsilon}}$.  Let us emphasize that in the axon we deal with the cord of small diameter, $2r$, and this thin cord  is wrapped by the myelin sheath of the length $2a$ per each segment, but for the model we consider electrolyte sphere with the radius $a$. Thus the auxiliary  concentration, $ n$,  of ions in the fictitious sphere  corresponds in fact to the concentration of ions in the cord, $n'=\frac{n 4/3 \pi a^3}{2a \pi r^2}$, which gives the typical cell concentration of ions $n' \sim 10 $ mM (i.e., ca. $6\times 10^{24}$ 1/m$^3$). This is due to the fact that in the dipole oscillation  in the segment of length of $2a$ take part all ions despite as in the fictitious model sphere or, as in the  real system, comprised in fact  to the much smaller volume of the thin cord portion, whereas the insulating myelinated sheath is a lipid substance without any ions. This insulating, relatively thick myelin coverage creates the tunnel required for formation of plasmon-polariton and its propagation inside such a channel. To reduce a coupling with surrounding inter-cellular electrolyte and protect against a leakage  of the plasmon-polariton, the myelin sheath must be sufficiently thick, much thicker than only for electrical  isolating  role.  Moreover, to fit better with conductivity parameters we accounted for, that for highly prolate geometry of actually oscillating ionic system,  the Mie frequency of  the  longitudinal oscillations is significantly lower than that one for the sphere with diameter  equal to the longer axis (for a rough estimation we assumed the factor 0.1).

For this Mie frequency, $\omega_1\simeq 4 \times 10^6$ 1/s, one can determine plasmon polariton self-frequencies in the chain of spheres with radius 50 $\mu$m (for Schwann cell length of $2a$) and the small chain separation, $d/a=2.01, 2.1,2.2$ (giving the Ranvier node length, $0.5,\;10,\;20\;\mu$m, respectively) using approach presented in the previous Appendices. The derivative of the self-frequency with respect to the wave vector $k$ determines  the group velocity of plasmon-polariton modes. The results are collected in Fig. \ref{axon-100}. We notice  that for ion system parameters as listed above, the group velocity of the plasmon-polariton reaches 100 m/s, for longitudinal modes, and 40 m/s, for transversal modes. Neverheless, if one assumes a larger transversal plasmon oscillation frequency \cite{elipsa1,elipsa2} for more realistic model of prolate spheroids (as for the inner Shwann cell cord with ions), one arrives also with the value of 100 m/s for transversal group velocity of plasmon-polariton for only twice increase of the frequency  (cf. Fig. \ref{axon-100}). The transversal ion oscillations are, however, inconvenient in the considered structure and moreover, the initial action potential post-synaptic or from the synapse  hillock, excites rather the longitudinal oscillations. 
 
  \begin{figure}[h]
\centering
\scalebox{0.3}{\includegraphics{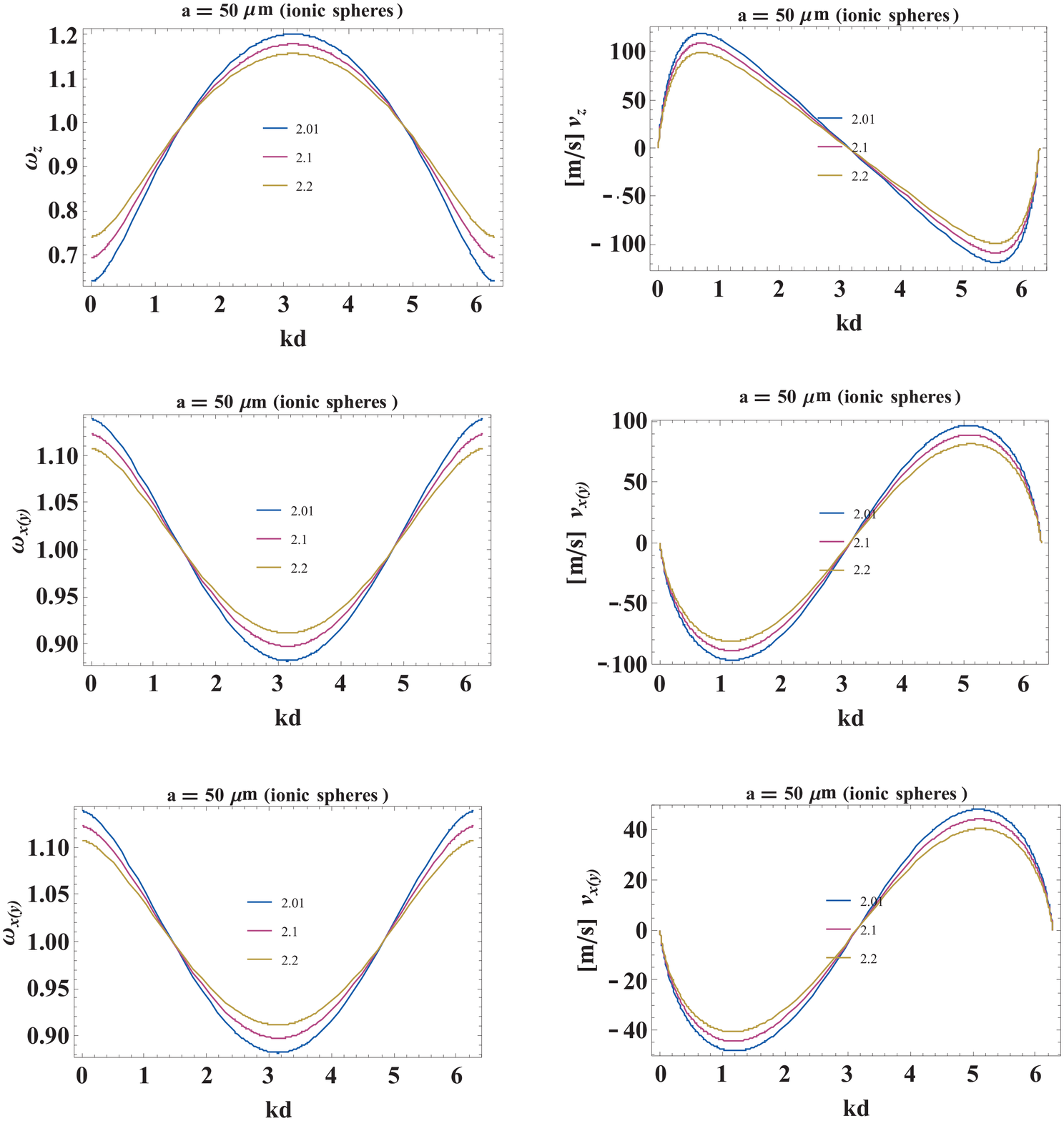}}
\caption{\label{axon-100} Solutions for the self-frequency and the group velocity of the  longitudinal (upper) and transversal (middle, lower) modes of plasmon-polariton in the model ionic chain; $\omega$ in $\omega_1$ units, here $\omega_1=4 \times 10^6 $ 1/s, for the chain of spheres with radius $a=50\;\mu$m and chain separation $d/a=1.01,1.1,1.2$, the equivalent ion concentration in the inner ionic cord of the axon, $n'\sim 6\times  10^{24}$ 1/m$^3$, the group velocity amplitude (for the longitudinal mode) $\simeq 100$ m/s---in the upper panel;  for the transversal mode the similar velocity amplitude is reached for twice-larger frequency ($2\times\omega_1$) assumed  in the prolate geometry \cite{elipsa1,elipsa2}; in the lower panel the transversal mode characteristics, but  for not increased  $\omega_1$  }
\end{figure}

Note, that for $\omega_1=4\times 10^6$ 1/s and $a=50 \;\mu$m the interference condition $kd-\omega_1 d/c=0$ and $kd+\omega_1 d/c=2\pi$ are fulfilled for extremely small values of $kd$ and $2\pi -kd$, respectively (of order of $10^{-6}$ for $d\in[2,2.5]$), thus are negligible for kinetics of plasmon-polariton at these conditions. Hence, the singularities caused by far- and medium-field contributions to dipole interaction are effectively removed by pushing to borders of the $k$ wave vector domain and moreover, practically whole the region $kd\in[0,2\pi)$ corresponds to ideal cancellation of the radiative losses of the plasmon-polariton modes. Additionally,  the mentioned singularities characteristic for infinite chains, cannot be also fully developed as the nerve chains are of finite length.

  \begin{figure}[h]
\centering
\scalebox{0.35}{\includegraphics{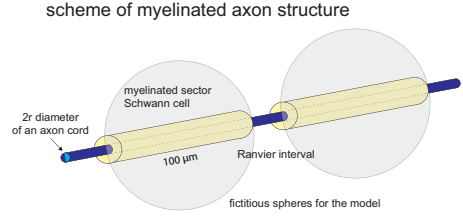}}
\caption{\label{axon-predkosc} Schematically presented periodic fragment of the myelinated long axon (myelin is a multilamellar lipid membrane which enwraps the axon cord in segments separated by nodes of Ranvier; the myelin sheath  is produced by special cells---Schwann cells  in the peripheral nervous system, and oligodendrocytes  in the central nervous system), for the model the fictitious periodic spherical ionic system chain is proposed  }
\end{figure}

Despite that the model of ionic system chain for myelinated axon is an  apparently crude approximation of the real axon structure, it can serve for comparison of energy and time scale of plasmon-polariton propagation in the model with kinetic parameters of nerve signals.  
In the model, the propagating in the axon chain plasmon-polariton, excited by the initial action potential on the first Ranvier node (after the  synapse or in the synapse hillock for the reverse signal direction), ignites  one-by-one the consecutive  Ranvier node blocks of $Na^+$, $K^+$ ion gates and the resulting fire of action potential traverses the axon with velocity of ca. 100 m/s actually observed in axons. Each new ignition releases energy supplement at particular Ranvier node block (due to its nonlinearity  the signal growth  saturates on the constant level and  the overall timing of each action potential spike has  stable shape  of local polarization/depolarization scheme of the short fragment of cell membrane corresponding to the Ranvier node). This permanent energy income contributes the plasmon-polariton dynamics compensating Ohmic thermal losses and assures  undamped its propagation over unlimited range. Though the whole signal cycle of action potential on a single Ranvier node block takes a time of several miliseconds (or even longer including the time for restoring the steady conditions, which, on the other hand, blocks reversion of the signal), the subsequent nodes are ignited earlier, accommodated to the velocity of the triggering pasmon-polariton wave packet. The velocity direction of the wave packet of  plasmon-polariton is adjusted to the semi-infinite chain geometry (actually the finite chain and excited at its beginning).  The action potential firing triggered  by plasmon-polariton travels along the axon in one direction also because the already fired nodes have  $Na^+$ gates  inactivated and need relatively long time to restore its original status (whole the block sodium/potassium needs some time, of second order, and energy supply to bring their concentrations to the normal values by across-membrane  active ion pumps). The plasmon-polariton scheme of ignition of action potential spikes in the chain of Ranvier nodes along the axon fits well with observed saltatory conduction of myelinated axons. 

  \begin{figure}[h]
\centering
\scalebox{0.35}{\includegraphics{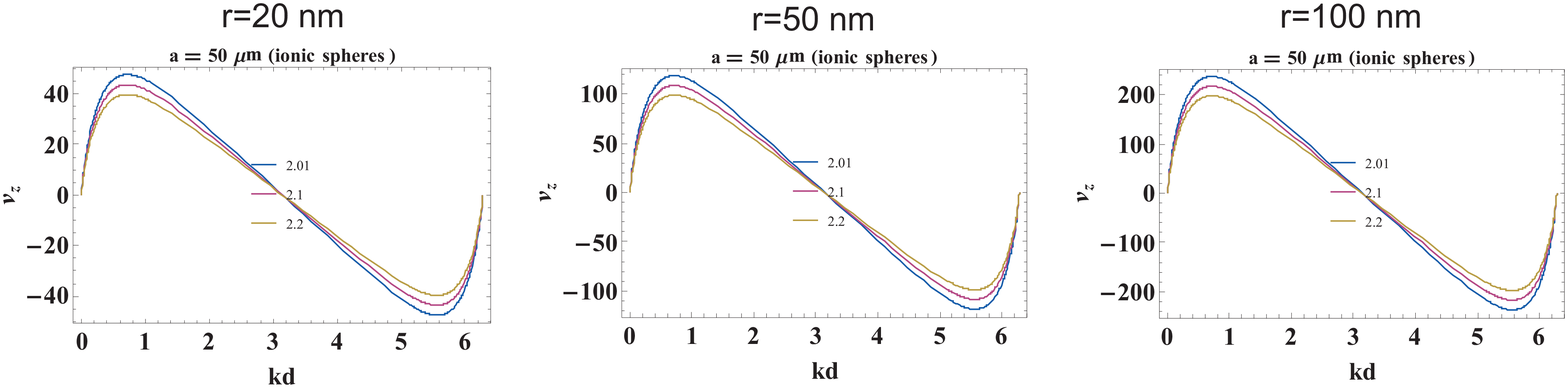}}
\caption{\label{axon-predkosc} Comparison of the group velocity, in [m/s], for the longitudinal plasmon-polariton mode with respect to the wave vector $k\in[0,2\pi/d)$, in the axon model with the Schwann cells of the  length 100 $\mu$m, 
Ranvier separation 0.5 $\mu$m, 5 $\mu$m, 10 $\mu$m (idicated by $d=2.01,\;2.1,\; 2.2$ in the figure) and radius $r=20,\;50,\;100$ nm of the axon cord}
\end{figure}

In Fig. \ref{axon-predkosc} the group velocity for action signal traversing in firing myelinated axon is compared for rising diameter of internal cord of the axon (cf. Fig. \ref{axon-predkosc}) (for concentration of ions in the cord $8$ mM (i.e., $\sim 3\times 10^{24}$ 1/m$^3$, typical for cell ion concentration), $100\;\mu$m length of myelinated sectors wrapped with Schwann cells and Ranvier interval of length 0.5 $\mu$m, 5 $\mu$m, 10 $\mu$m. The dependence on the Ranvier interval length is weak (not important) but the rise of the velocity with the thickness  
of the internal cord is significant.

\end{document}